# Simulation Platform for Wireless Sensor Networks Based on Impulse Radio Ultra Wide Band

*Abdoulaye Berthe, Aubin Lecointre, Daniela Dragomirescu, and Robert Plana*
University of Toulouse
LAAS - CNRS
7, Av du Colonel Roche
31077 Toulouse Cedex 4
{aberthe, alecoint, daniela, plana} @laas.fr

**Abstract**

*Impulse Radio Ultra Wide Band (IR-UWB) is a promising technology to address Wireless Sensor Network (WSN) constraints. However, existing network simulation tools do not provide a complete WSN simulation architecture, with the IR-UWB specificities at the PHYsical (PHY) and the Medium Access Control (MAC) layers. In this paper, we propose a WSN simulation architecture based on the IR-UWB technique. At the PHY layer, we take into account the pulse collision by dealing with the pulse propagation delay. We also modelled MAC protocols specific to IR-UWB, for WSN applications. To completely fit the WSN simulation requirements, we propose a generic and reusable sensor and sensing channel model. Most of the WSN application performances can be evaluated thanks to the proposed simulation architecture. The proposed models are implemented on a scalable and well known network simulator: Global Mobile Information System Simulator (GloMoSim). However, they can be reused for all other packet based simulation platforms.*

## 1. Introduction

A WSN consists of several sensor nodes scattered in a small area [1-2]. The role of these nodes is to sense a particular phenomenon and to report it to a base station, for analysis. Such networks can be used for applications like monitoring, local area control, industrial processes, civil safety, house automation and tactical applications [1-2]. However, we have to deal with several constraints when setting up a large scale autonomous WSN: power consumption, simplicity, low cost, small size [3]. Existing solutions based on the Wireless Local Area Networks (WLAN) or the Wireless Personal Area Networks (WPAN) standards do not address all the mentioned issues. One of the promising solutions which can address them is the Time Hopping IR-UWB (TH-IR-UWB) technique [4]. Its main advantages are low power consumption, low collision probability thanks to the time hopping technique and short pulse duration [5]. TH-IR-UWB is a kind of dynamic Time Division Multiple Access (TDMA) at pulse level [6]. The medium is divided into frames and frames are divided into time slots. The nodes transmit one pulse per frame and the pulse position on a frame depends on the transmitter Time Hopping Sequence (THS). Thanks to this technique, TH-IR-UWB offers multiuser access capabilities. Unfortunately, available network simulation tools [7] suffer from a lack of sensors, sensing channels and accurate TH-IR-UWB models at the PHY and MAC layers. This work presents a generic WSN simulation platform that can be used to evaluate the whole WSN architecture based on TH-IR-UWB, following criteria like energy efficiency, latency or reliability. The remainder of this paper is organized as follows. In Section 2 we describe the problem. Section 3 presents related work. The proposed PHY and MAC layer models specific to TH-IR-UWB are described respectively in Section 4 and 5. Section 6 presents sensor and sensing channel models and finally, Section 7 concludes.

## 2. Problem formation

Our concern is the PHY layer which is in charge of coding and modulation, and the MAC layer which controls the medium access [8]. Our goal is to model the TH-IR-UWB technique at these layers in order to analyze its impact on the whole network architecture, in the WSN context. For this purpose, four main key factors have to be considered: the transceiver performance, the error rate, the energy consumption problematic, the pulse collision and the channel access scheme. For each of them, the TH-IR-UWB specificities must be considered, especially the discontinuous aspect.

## 2.1. PHYsical layer

The main motivation of a PHY layer model is to accurately represent the signal-to-bit and bit-to-signal conversion. The best way to do it is to use a Bit Error Rate/Signal to Noise Ratio (BER/SNR) model with a network simulator. This technique allows a good balance between simulation time and accuracy. In contrast, another solution which consists to an interaction between a network simulator and a digital signal processing tool is not viable due to the large amount of processing time, especially in the WSN context, where we have to deal with thousands of nodes. We can obtain a BER/SNR table either by simulation with digital signal processing tools such as Matlab [9], or by measuring the BER on real transceivers. This table is used at the network simulator PHY layer to evaluate and simulate the Packet Error Rate (PER), in order to determine if a packet has to be delivered to the MAC layer or not. In addition to this concept, we have to deal with the interference or collision management, which consists to detect overlapping transmissions at a particular receiver. Together, these two schemes lead to a high fidelity in the network simulation, especially with single channel techniques where Continuous Wave (CW) is used. However, with techniques like TH-IR-UWB we must pay particular attention to the collision model, since the transmitted signal is in form of very short impulses instead of CW. The channel does not appear to be occupied during the whole packet transmission duration. So, two or more concurrent transmissions do not systematically overlap. We should also notice that the TH-IR-UWB technique is different to FH (Frequency Hopping), where transmission performed with perfect orthogonal sequences does not interfere. With TH-IR-UWB, the orthogonality has to be considered at a particular receiver, as the pulse reception time depends on the pulse propagation delay. In our proposed model, we try to represent the real behaviour of the PHY layer. Interference management is performed at the receiver and at the pulse level, by taking into account the pulse propagation delay.

## 2.2. MAC layer

Due to the impulse behaviour of TH-IR-UWB, traditional MAC protocols, for WLAN, or WPAN, which are generally based on the carrier sensing mechanism seem to be inefficient. In fact, the free channel concept changes and becomes difficult to define. Moreover, the Request to Send/Clear to send (RTS/CTS) also called virtual carrier sensing and the TDMA schemes contribute to reinforce mutual exclusion in the channel access. Thus, they do not benefit from the multiuser access capabilities offered by TH-IR-UWB. In addition, carrier sensing and virtual carrier sensing techniques lead to high power consumption. By keeping in mind these facts and the simplicity constraint, we also model MAC protocols for TH-IR-UWB.

## 3. Related work

In the literature, several TH-IR-UWB MAC and PHY layer models have been proposed [10-13]. However [11-13] do not provide a network simulation architecture. In [11], a MAC layer protocol (UWB)[2], which takes into account the TH-IR-UWB PHY layer benefits is proposed. A common channel, defined by a particular THS is used to exchange the THS before the transmission. In [12], a probabilistic pulse collision model is proposed while [13] investigates on finding appropriate THS in multiuser access networks. We lead our research in a context similar to [10], where a wireless network simulation architecture based on TH-IR-UWB has been proposed and realized over the Network Simulator-2 (NS-2) [14]. The proposed model concerns the PHY and MAC layers. At the MAC layer, it addresses the issues of power control. At the PHY layer, it uses a BER/SNR channel model and takes into account the sub channel division, thanks to the THS. The interference management is performed according to the average orthogonality between two transmissions using the same THS and the average orthogonality between two transmissions, using different THS [15]. However, these parameters being computed in advance, the impact of the pulses real propagation delay disappears on the network simulator PHY layer model. In addition, this model is not directly addressed to WSN. It does not provide sensor and sensing channel models. Its probabilistic choice of the successful communication among the interfering ones, which may not always represents the real PHY layer behaviour, leads to a lack of accuracy. We should also notice that transmission, even if performed with perfect orthogonal THS can collide at a particular receiver, according to the pulse propagation delay. This fact is not accurately taken into account in the models in literature.

## 4. Proposed PHYsical layer model

The ultimate goal of our model is to accurately detect pulses that may interfere at a particular receiver. Moreover, we take into account the BER/SNR [6], in order to reflect some TH-IR-UWB coding, modulation schemes or transceivers performance. With TH-IR-

UWB, a transmitter sends pulses following a THS. A pulse transmitted at the time $t$ is received at a particular receiver after the pulse propagation delay; at $t + \Delta t$. $\Delta t$ depends on the distance $d$ between the transmitter and the receiver and also the pulse velocity $v_{pulse}$.

$$\Delta t = \frac{d}{v_{pulse}} \quad (1)$$

The velocity in turn depends on the center of the occupied bandwidth $f_c$ and the wavelength $\lambda$.

$$v_{pulse} = \lambda f_c \quad (2)$$

From the above formulations, we can deduce that the pulse reception time at a particular receiver is not constant. Indeed, it depends on the distance between the two considered nodes, and the central frequency of the UWB signal. As a result, the pulse collision at a particular receiver depends on the transmission THS and also the propagation delay. These facts lead to the lack of precision of models that do not take into account the real pulse propagation delay like [10] since the pulse collisions depends on it. They also demonstrate that the use of perfect THS does not systematically avoid pulse collisions or interferences, even in case of perfect synchronization.

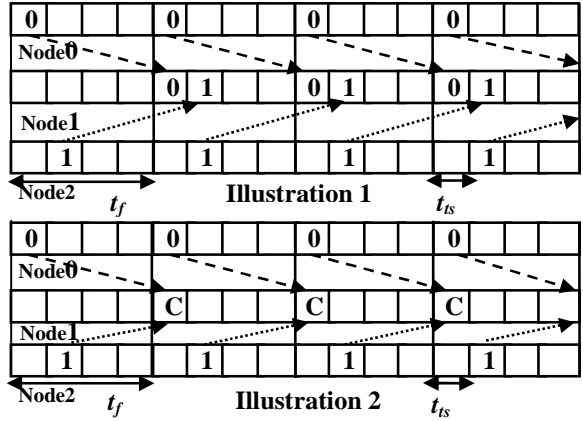

**Figure 1: Illustration of collisions.**

In the Figure 1, three nodes are perfectly synchronized. Transmissions are performed with orthogonal THS but they overlap at the receiver. Node1 and Node2 are transmitting a sequence of pulses to Node0 in the time slot THS1 equals to 1 for Node1 and the time slot THS2 equals to 2 for Node2. Let us denote $\delta_1$ the pulse propagation delay between Node1 and Node0 and $\delta_2$ the pulse propagation delay between Node2 and Node0. In the Illustration 1, pulse collision does not occur because $\delta_1$ equals to $\delta_2$ and the two sequences (1111) and (2222) are perfectly orthogonal. In Illustration 2, however, the transmitted pulses collide in spite of the use of the previous sequences by the two transmitters. This is due to the pulse propagation delay. In case where the THS length is equal to one, like in this example, collisions occur when:

$$\delta_1 = \delta_2 \text{ and } \neg(THS_1 \perp THS_2) \quad (3)$$

Or

$$(THS_1 \perp THS_2) \text{ and } \delta_2 = (\delta_1 - (THS_1 - THS_2) t_{ts}) \quad (4)$$

Or

$$(THS_1 \perp THS_2) \text{ and } \delta_2 = (\delta_1 - (THS_2 - THS_1) t_{ts}) \quad (5)$$

In these formulas, $t_{ts}$ denotes the time slot duration.

### 4.1. Hypothesis

The underlined facts above constitute the backgrounds of our model. To set it up, we consider the TH-IR-UWB principle [6] at each transceiver: frame and slot division. For each defined time slot at a particular transceiver, we consider four states:

- IDLE: There is neither transmission nor reception in this time slot.
- TRANSMIT: The PHY layer is transmitting a pulse in this time slot.
- SLEEP: The PHY layer does not listen in this time slot.
- SENSE: The PHY layer is receiving a signal whose power is too low be decoded in this time slot.
- RECEIVE: The PHY layer is receiving a pulse on this time slot.

The defined states above depend on the received or transmitted pulse power. Thus, we accurately define the noise level at each receiver.

$$P_n = N_0 W \quad (6)$$

Where $W$ denotes the occupied bandwidth, $N_0 = KT$ is the thermal noise density. $K$ denotes the Boltzmann constant and $T$ is the noise temperature. This is important because the noise power depends on the occupied bandwidth. Furthermore, in a BER/SNR model, the probability of good reception depends on the received $SNR$. When two pulses arrive on the same time slot at a receiver, a collision occurs and the pulse which has the lowest level is added to the noise power. We perform the interference management according to the Signal and Interference to Noise Ratio (SINR) computation [9], by taking into account the noise figure F.

$$SINR = \frac{P_S}{\sum_{all\ other} + FP_n} \quad (7)$$

Where $P_s$ is the receiving pulse power and $F$ denotes the noise figure. $\Sigma$ *all other* denotes all the interfering signals where the received power of each of them is lower than $P_s$. They are supposed to be conformed to *Gaussian noise*. In addition, we made the following important assumptions:

- The propagation delay is supposed to be constant during a packet transmission.
- To simplify THS management we consider multi user networks where the THS length is equal to one.
- The PHY layer transmits one packet a time.
- The PHY layer is able to receive concurrent transmissions only if they do not interfere with each other.
- Each started transmission must go on until all the pulses that constitute the packet are sent.
- Nodes are supposed to be synchronized on the TH-IR-UWB frames.

### 4.2. Representation

At the network simulator PHY layer, we represent our impulse radio by a set of time slots according to the TH-IR-UWB principle. The time slot duration is $t_{ts}$ and its representation can include the delay spread. In this case, overlapping reception detection is performed on a duration which includes this delay. Thanks to the nanosecond simulation accuracy offered by discrete event simulation and the synchronization hypothesis, we are able to deduce the current time slot at a particular PHY layer by only considering the simulation clock value. This is possible because the frame repetition is a periodical process and time slot position within the frames does not change.

$$t_{cur} = E\left\lfloor\frac{t_{sim} \bmod t_f}{t_{ts}}\right\rfloor \quad (8)$$

Where $E\lfloor X \rfloor$ denotes the floor function, $t_{cur}$ gives us the position of the current slot, $t_f$ denotes the frame duration and $t_{sim}$ is the simulation clock value at the pulse reception.

### 4.3. Transmission

A packet transmission process in TH-IR-UWB consists in transmitting a pulse on a slot in each frame until the full packet is transmitted. We model this process in the network simulator by modifying the state of the affected slots according to the transmitter THS. Transmission always starts on the first time slot of the sequence and lasts $T_{trans}$.

$$T_{trans} = l_{pdu} t_f \quad (9)$$

Where $t_f$ denotes the frame duration and $l_{pdu}$ denotes the packet length in bits. We only modify one slot state as the transmitter always transmits on the same time slot for duration $T_{trans}$. As illustrated in Figure 2, the wait function determines the position of the current time slot and delays transmission until the next access time slot position.

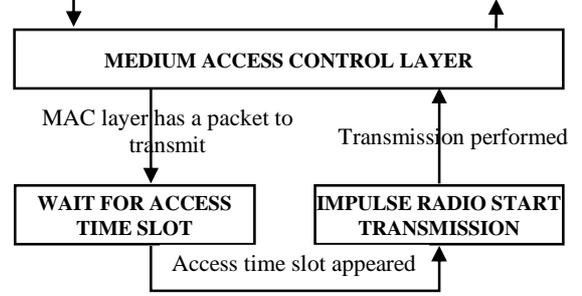

**Figure 2: Transmission process.**

### 4.4. Reception

The packet reception model is a main feature in a PHY layer model. We deal with the reception process by considering slots states at the receiver.

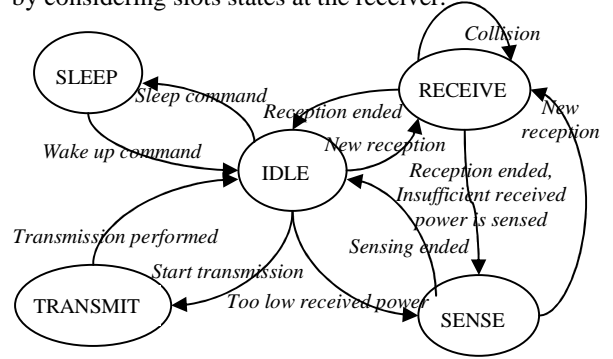

**Figure 3: Slot state transition.**

When the PHY layer detects a new packet on the medium, we first determine the position of the occupied time slot (8). The reception sequence of a particular receiver depends on the THS used for the transmission and the propagation delay. In our model, as the THS length is supposed to be one, we only consider one time slot state. The transmitter sends the pulses on the same slot all along during the packet transmission. According to the reception slot state, different transitions are possible as illustrated in Figure3

### 4.5. Power consumption

The proposed power consumption model is an enhanced GloMoSim power consumption model [16]. Furthermore it takes into account the impulse nature of TH-IR-UWB. The global power consumption computation is based on the power consumed per pulse transmission or reception.

## 5. Proposed MAC layer models

By taking into account the facts outlined in Section 2, we modelled two MAC protocols: Unslotted-TH and Slotted-TH and their reliable versions: Reliable-Unslotted-TH and Reliable-Slotted-TH. These TH-IR-UWB MAC protocols are very simple thanks to the use of TH at the PHY layer. Indeed, TH randomizes the pulse positions, thus coupled with very short pulse duration, the pulse collision probability decreases. As a result, the MAC layer of this kind of technique can be very simple. Thus, complex and power hungry mechanisms such as carrier sense and channel reservation have no longer reasons to be used.

### 5.1. Unslotted-TH

In Unslotted-TH, the transmitter does not care about the channel state. Like ALOHA [17], once it has a packet to send, it transmits it on the medium, according to its own THS.

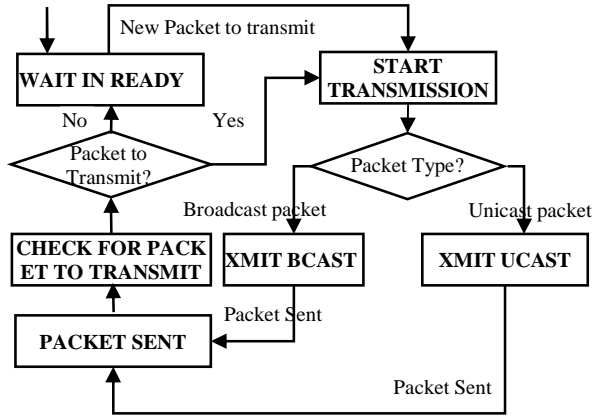

**Figure 4: Unslotted-TH illustration.**

As the received packets are not acknowledged here, no retransmission is needed. This protocol leads to low latency and gives a high priority to new events to be notified to the base station in a WSN application. It well suits applications where latency and new events notification are critical. It is also accurate for power management in a *piconet* structure. Nodes can easily be put in doze mode when no communication has to be performed. However, this becomes more complex in a large scale WSN where routing is also performed by sensor nodes. The different states of Unslotted-TH and the transitions between them are illustrated in Figure 4.

### 5.2. Slotted-TH

In Slotted-TH, the medium is shared in MAC frames and the transmitter is only allowed to start a transmission at the beginning of a MAC frame, without carrier sensing. Medium sharing at the MAC layer introduces the problem of access scheduling. It can be performed in a distributed or centralized manner. And we can reinforce mutual exclusion by allocating non overlapping MAC frames to different nodes or allow it to benefit from the multiuser access capabilities offered by the TH-IR-UWB. In this protocol, the slot size impacts the latency. And, like Unslotted-TH, the energy management can be easily performed; medium sharing into MAC frames makes this process simpler. In case of centralized slots allocation, the coordinator choice problem occurs with large scale autonomous WSN. But, if the sensor nodes positions are known in advance, it can be easily performed. Figure 5 exhibits the Slotted-TH principle.

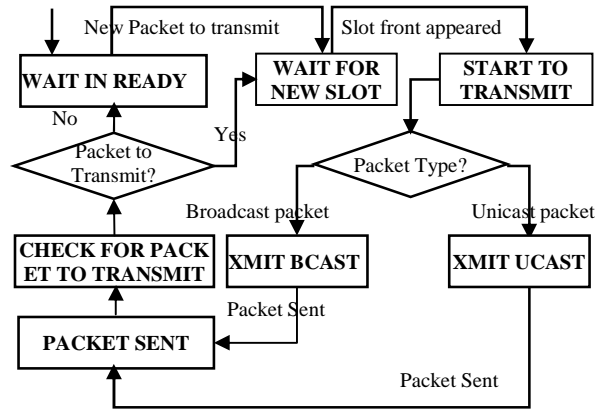

**Figure 5: Slotted-TH illustration**

### 5.3. Reliable-Unslotted-TH

Reliable-Unslotted-TH is the reliable version of Unslotted-TH. It is very similar to ALOHA and requires received packet acknowledgment. After sending a packet, the transmitter, waits for a retransmission delay. If it receives an acknowledgment before this delay expiration, it transmits a new packet; otherwise it retransmits the current packet until the number of retransmission exceeds the retransmission limit. This protocol is adapted to WSN applications where reliability is critical. By increasing the retransmission limit we increase reliability but this may also impact latency. By increasing the retransmission delay we increase latency. Due to the packet acknowledgment process, power management become more difficult to be performed. However, nodes can be put in doze mode after the current packet acknowledgment reception.

### 5.4. Reliable-Slotted-TH

Reliable-Slotted-TH is the reliable version of Slotted-TH. The retransmission delay becomes the delay until the next slot front appears. The retransmission limit and the retransmission delay respectively impact reliability and latency. Here, the energy management problem is similar to Reliable-Unslotted-TH.

## 6. Sensor and sensing channel

Sensor and sensing channel models are key features in a WSN simulation. However most of the packet based network simulators do not provide them. We propose here a generic and reusable model. To achieve an accurate representation of the sensing sub system, we consider the inherent characteristics of the sensing device and the phenomenon to be sensed. Furthermore, we derived some important features: the sampling rate, the sensing range and the device performances (false negative and false positive). In practice, most of the phenomena to be sensed (seismic vibration, acoustic wave etc.) can be represented by wave propagation. They can be emulated as an electromagnetic propagation at a particular frequency, quite similar to a well modelled PHY layer of some network simulator. We model the phenomenon to be sensed as a periodic broadcast of an electromagnetic wave at the sampling rate of the sensor device. The simulator determines according to the distance and the sensing channel path loss, the received intensity at the sensor side. Some thresholds are defined, in order to accurately reflect the performances of the sensing devices. They can be tuned to meet real sensors behaviour. This model, in contrast to the proposed model in [18-19] is not dedicated to a particular phenomenon, so it can be easily reused and many types of sensor devices can be modelled with this generic technique.

## 7. Conclusion and future work

In this paper we presented a WSN simulation architecture based on IR-UWB. At the PHY layer we propose a model that takes into account pulse collision. At the MAC layer, we model MAC protocol for WSN applications by taking into account the TH-IR-UWB specificities. We also propose a generic and reusable sensor and sensing channel models based on electromagnetic wave propagation. The presented architecture is modular so it can be easily modified and reused.